\documentclass[mathleft,fleqn,%
]{an}

\usepackage[varg]{txfonts}
\usepackage{natbib}
\usepackage{authblk}
\usepackage{graphicx}     
\usepackage{amssymb}    
\usepackage{multicol}      
\usepackage{bm}             
\usepackage{float}           
\usepackage{tipa}            
\usepackage{wasysym}
\usepackage{color,soul}

\begin{document}

\title{Interacting Fields and Flows: Magnetic Hot Jupiters}
\author{Daley-Yates S.\inst{1}\fnmsep\thanks{Corresponding author:
        {sdaley@star.sr.bham.ac.uk}}
\and  Stevens I. R.\inst{1}
}
\titlerunning{Magnetic Hot Jupiters}
\authorrunning{Daley-Yates S. \& Stevens I. R.}
\institute{
School of Physics and Astronomy, University of Birmingham, Edgbaston, Birmingham, B15 2TT}

\received{XXXX}
\accepted{XXXX}
\publonline{XXXX}


\abstract{We present Magnetohydrodynamic (MHD) simulations of the magnetic interactions between a solar type star and short period hot Jupiter exoplanets, using the publicly available MHD code PLUTO. It has been predicted that emission due to magnetic interactions such as the electron cyclotron maser instability (ECMI) will be observable. In our simulations, a planetary outflow, due to UV evaporation of the exoplanets atmosphere, results in the build-up of circumplanetary material. We predict the ECMI emission and determine that the emission is prevented from escaping from the system. This is due to the evaporated material leading to a high plasma frequency in the vicinity of the planet, which inhibits the ECMI process.}

\maketitle

\section{Introduction} 
\label{sec:intro}

Magnetic Star-Planet Interaction (mSPI) is an excellent candidate for generating observable radio emission from exoplanetary systems. Non-thermal cyclotron emission via the Electron Cyclotron Maser Instability (ECMI) is an efficient mechanism for producing such emission \citep{Zarka2007, Stevens2005a}. Incident solar wind power, both ram and magnetic, on a planetary magnetic field is converted to radio emission.

Radio emission in the form of the ECMI emission is linearly dependent on the local magnetic field strength. Observing this emission would therefore allow an indirect means of quantifying the magnetic field strength of exoplanets. This would constrain internal structure models and planetary rotation, informing us about the evolutionary history of the planet and host system \citep{Hess2011}. Due to the close proximity to their host stars and their potentially strong magnetic fields, hot Jupiters present the best candidates for exoplanet radio emission.

Such emission is thought to be detectable. However, to date, there have been no confirmed radio observations of exoplanets. This lack of detection invites explanation via theoretical studies.

Numerous numerical studies have been carried out in recent years which have simulated many aspects of SPI, examples range from detailed studies of the close in atmospheres of hot Jupiters \citep{Strugarek2014b, Khodachenko2015, Vidotto2015}, the atmospheres of their host stars \citep{Alvarado-Gomez2016, Alvarado-Gomez2016a}, to the global structure of the planetary-stellar wind interaction \citep{Bourrier2013, Bourrier2016, Owen2014, Alexander2016}. The study presented here builds on this work, specifically on the simulations conducted by \cite{Matsakos2015}, extending it to investigate ECMI emission in the context of mSPI. 

\section{Modelling}    
\label{sec:modelling}

Using the approach of \cite{Matsakos2015} we have constructed two model exoplanets, which differ only with respect to their mass-loss rates. The planetary parameters for the models are summarized in Table \ref{tab:params}. These values are constant with simulations by \cite{Salz2016} for hot Jupiters such as WASP-12 b \citep{Hebb2009} and GJ 3470 b \citep{Bonfils2012}. 

The Parker model of the solar wind is used to initialise the simulation \citep{Parker1958}. The equation:
\begin{equation}
\label{eq:parker_eq}
	\frac{v^{\mathrm{init}}_{\mathrm{W}} (r)^{2} }{c_{\mathrm{s}}^2} - \ln{ \left( \frac{v^{\mathrm{init}}_{\mathrm{W}} (r)^{2} }{c_{\mathrm{s}}^{2}}  \right)} = - 3 - 4 \ln{ \left( \frac{v_{esc}^{2}}{4 c_{\mathrm{s}}^{2}} \right)} 
	+ 4 \ln{ \left( \frac{r}{R} \right) } + \frac{ R v_{esc}^{2}} {r c_{\mathrm{s}}^{2}}
\end{equation}
is used to initialise the wind of both the star and the planet. In this respect, the winds of both bodies are considered to follow the same physical processes, a thermally driven thermodynamic expansion. $r = \sqrt{x^{2} + y^{2} + z^{2}}$, with $x$, $y$ and $z$ being the Cartesian coordinates. $R$ is the radius of the body, $v_{esc}~=~\sqrt{2 G M/R}$,  the escape velocity and $c_{\mathrm{s}}~=~\sqrt{2 k_{\mathrm{B}} T / m_{\mathrm{p}}}$ the sound speed, with $k_{\mathrm{B}}$ the Boltzmann constant, $T$ the temperature and $m_{\mathrm{p}}$ the proton mass. The wind velocity, $v^{\mathrm{init}}_{\mathrm{W}}(r)$, is found by numerically solving equation (\ref{eq:parker_eq}), using a root finding algorithm. 

The stellar and planetary magnetic fields are initially dipolar, according to:
\begin{equation}
\label{eq:b_field}
    \bm{B}^{init} (x, y, z) = \frac{B_{eq} R^{3}}{r^{5}} 
    \left[ 3 x z \hat{x} + 3 y z \hat{y} + 
    \left( 3z^2 + r^2 \right) \hat{z} \right] 
\end{equation}
where $B_{eq}$ is the equatorial magnetic field of either the star or the planet. Everywhere external to the two bodies, the total field is the sum of the fields from the two bodies. Resulting in a field which is free from discontinuities. 

Stellar and planetary quantities used in the simulations are listed in Table \ref{tab:params}. Values were chosen which correspond to hydrostatic models described in \cite{Matsakos2015}. A solar strength magnetic field was chosen for the star. While the the planetary magnetic field strengths were chosen based on the assumption that the field is driven by dynamo action due to the rotation of the planet \citep{Stevenson2003, Strugarek2014b}. Since hot Jupiters are assumed to be tidally locked \citep{Griessmeier2004}, their rotation is much slower than that of Jupiter, which has an $B_{\mathrm{eq}} ≈ 15 \ \mathrm{G}$. As such a $B_{\mathrm{eq}} = 1 \ \mathrm{G}$ for the planet was used in both models. This value is consistent with those found in the literature \citep{Pillitteri2015, Strugarek2015, Nichols2016}.

\begin{table*}
\centering
\caption{Stellar and planetary parameters used in the simulations. The subscripts refer to either the star ($\ast$) or planet ($\circ$).  $M_{\ast, \circ}$ is the mass, $R_{\ast, \circ}$ the radius, $T_{\ast, \circ}$ the temperature, $B_{eq \ast, \circ}$ the equatorial magnetic field strength, $\rho_{\ast, \circ}$ the surface density, $a$ the orbital separation, $p_{\mathrm{orb}}$ the orbital period and $p_{\mathrm{rot} \ast, \circ}$ the rotational period.}
\begin{tabular}{ccccc}
\hline
Parameter & Star & Planet (model 1) & Planet (model 2)  \\
\hline
$M_{\ast, \circ}$ & $1 \ \mathrm{M_{\odot}}$ & $0.5 \ \mathrm{M_{J}}$ & $0.5 \ \mathrm{M_{J}}$ \\
$R_{\ast, \circ}$ & $1 \ \mathrm{R_{\odot}}$ & $1.5 \ \mathrm{R_{J}}$ & $1.5 \ \mathrm{R_{J}}$ \\
$T_{\ast, \circ}$ & $1 \times 10^{6} \ \mathrm{K}$ & $6 \times 10^{3} \ \mathrm{K}$ & $10^{4} \ \mathrm{K}$ \\
$B_{eq \ast, \circ}$ & $2 \ \mathrm{G}$ & $1\ \mathrm{G}$ & $1\ \mathrm{G}$ \\
$\rho_{\ast, \circ}$ & $5 \times 10^{-15} \ \mathrm{g/cm}^{3}$ & $7 \times 10^{-17} \ \mathrm{g/cm}^{3}$ & $7 \times 10^{-16} \ \mathrm{g/cm}^{3}$ \\
$a$ & $-$ & $0.047 \ \mathrm{au}$ & $0.047 \ \mathrm{au}$ \\
$p_{\mathrm{orb}}$ & $-$ & $3.7 \mathrm{days}$ & $3.7 \ \mathrm{days}$ \\
$p_{\mathrm{rot} \ast, \circ}$ & $3.7 \ \mathrm{days}$ & $3.7 \ \mathrm{days}$ & $3.7 \ \mathrm{days}$ \\
\hline
\end{tabular}
\label{tab:params}
\end{table*}

\subsection{Numerical Scheme}
\label{sec:Num}

The public code PLUTO (version 4.2) was used to solve the MHD equations \citep{Mignone2007, Mignone2012}. A second order scheme with linear reconstruction, a Runga-Kutta integrator, HLLD Riemann solver and the GLM solenoidal constraint method was employed.

The simulations covered an extent of $-32 \ R_{\ast}~<~x, y~<~32 \ R_{\ast}$ and $-16 \ R_{\ast}~<~z~<~16 \ R_{\ast}$ which was divided into a mesh with an initial resolution of $128^2$ cells in the $x$-$y$ plane and 64 cells in the $z$ direction. Adaptive mesh refinement was used to achieve an effective resolution of $4096^2 \times 2048$. The simulation was evolved through a total of $360 \ \mathrm{k s}$ in order to reach quasi-steady state.

\subsection{ECMI Emission}
\label{sec:ECMI}

The efficiency of ECMI emission is a balance between the cyclotron frequency, as a function of the magnetic field strength and plasma frequency, as a function of the local plasma density at the site of emission. The cyclotron emission in MHz is given by:
\begin{equation}
\label{eq:ce}
   \nu_{\mathrm{ce}} (\mathrm{MHz}) = \left( \frac{e B}{2 \pi m_{\mathrm{e}} c} 
   \right) = 2.80 B, 
\end{equation}
with $e$ and $m_{\mathrm{e}}$ the electron charge and mass, $c$ the speed of light and $B$ the local magnetic field strength. This is counteracted by the plasma frequency:
\begin{equation}
\label{eq:pe}
   \nu_{\mathrm{pe}} (\mathrm{MHz}) = \left( \frac{n_{\mathrm{e}} e^{2}}{\pi 
   m_{\mathrm{e}}} \right)^{1/2} = 8.98 \times 10^{-3} n_{\mathrm{e}}^{1/2},
\end{equation}
which acts to inhibit the cyclotron emission if
\begin{equation}
\label{eq:ECMI}
  \frac{\nu_{\mathrm{ce}}}{\nu_{\mathrm{pe}}} > 1.
\end{equation}
$n_{\mathrm{i}}$ is the ion number density. The right hand sides of equations (\ref{eq:ce}) and (\ref{eq:pe}) reduce down to constants, such that the resultant units are in MHz. For the emission process to be efficient, this ratio needs to be $\gtrsim 2.5$ \citep{Weber2017}. In the following section, this ratio for the circumplanetary material of both models and discussed in the context of observational emission will be shown.

\section{Results}
\label{sec:Res}

\begin{figure*}
\centering
\includegraphics[width=1\textwidth]{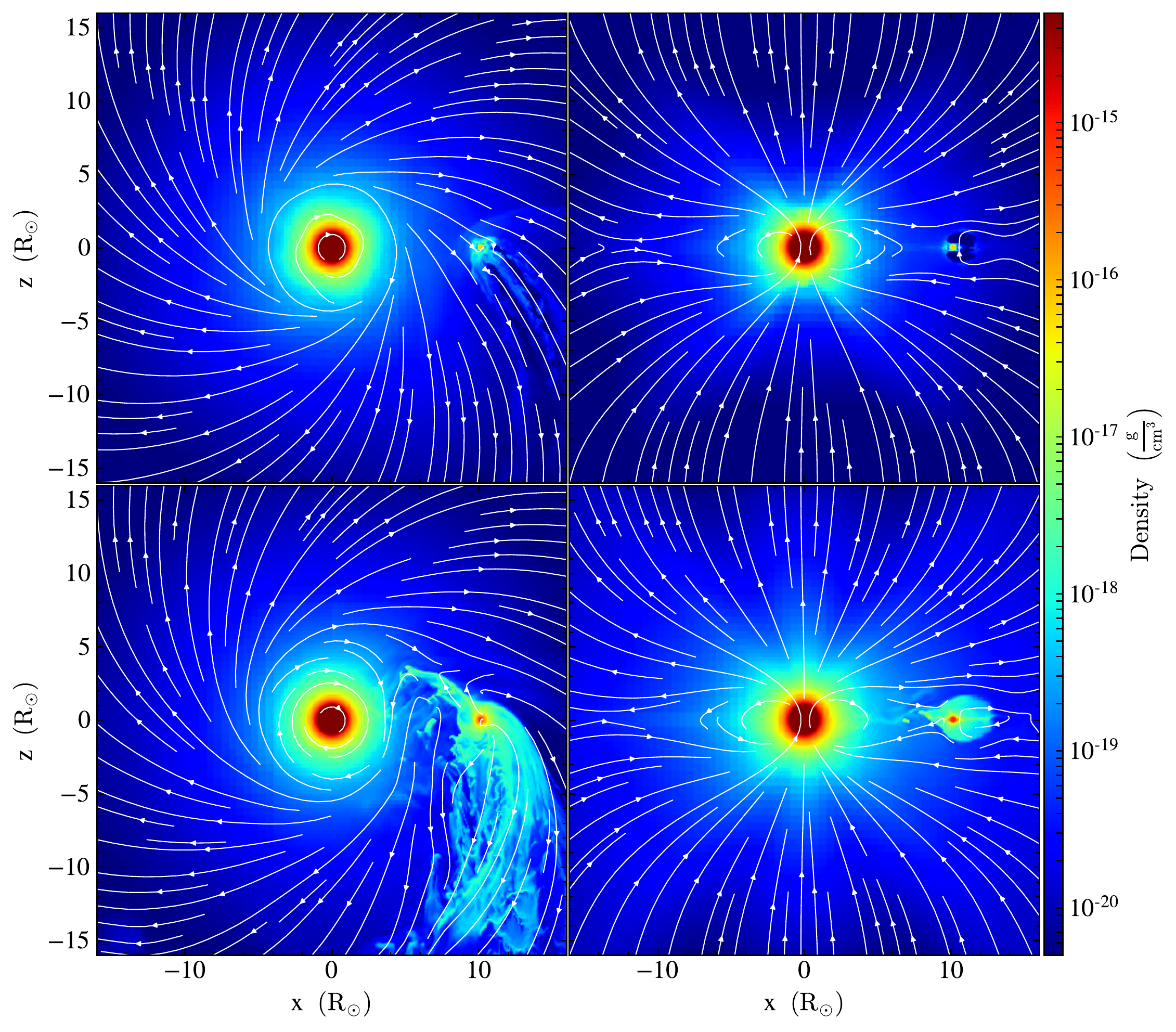}
\caption{Density colour plots of a portion of the simulation domain. The top row shows Model 1 and the bottom row shows Model 2. The left column shows the top down view on the simulations, with velocity flow lines. The right column shows the side on view with magnetic field streamlines.
\label{fig:global}}
\end{figure*}

\begin{figure*}
\centering
\includegraphics[width=0.7\textwidth]{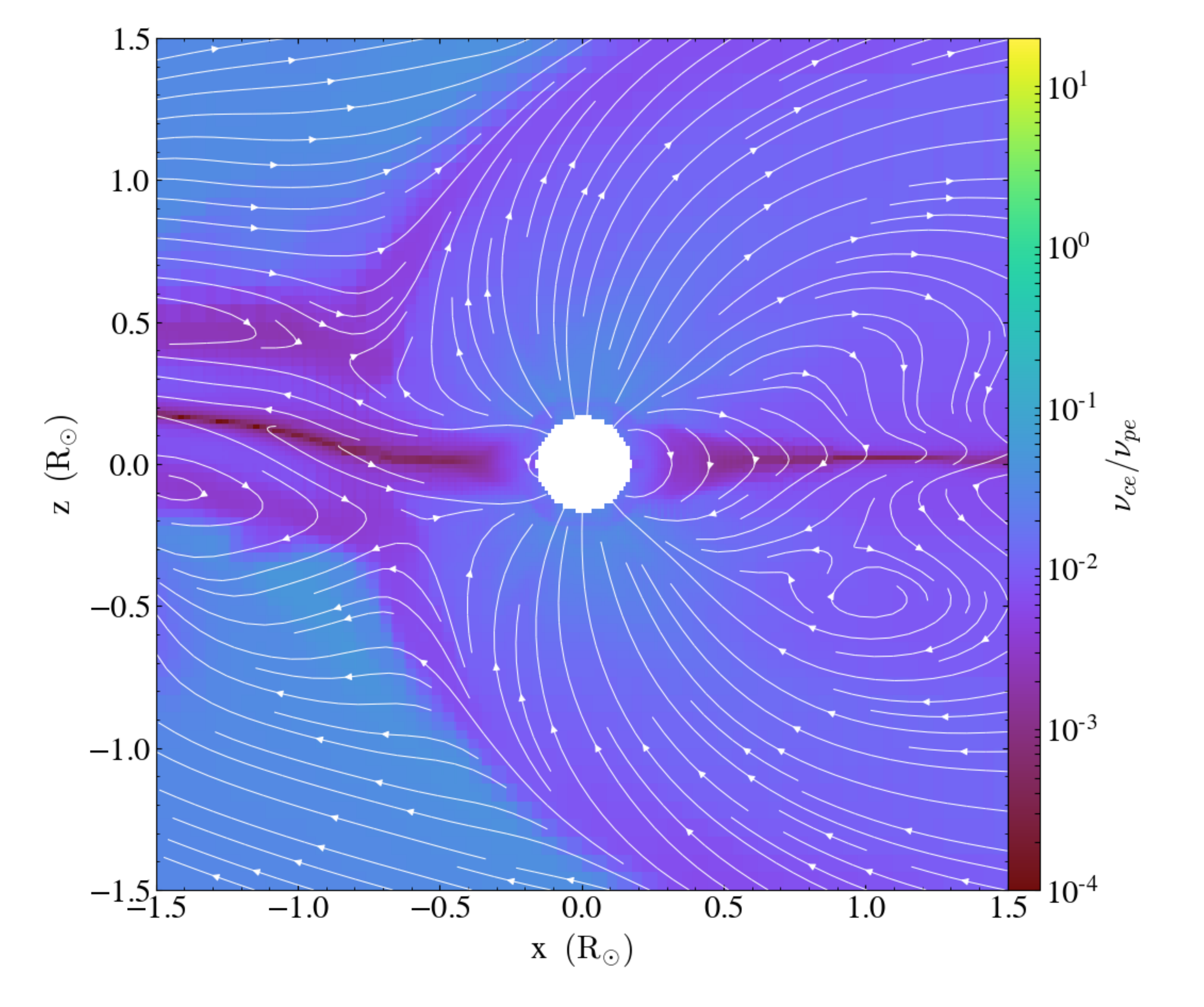}
\includegraphics[width=0.7\textwidth]{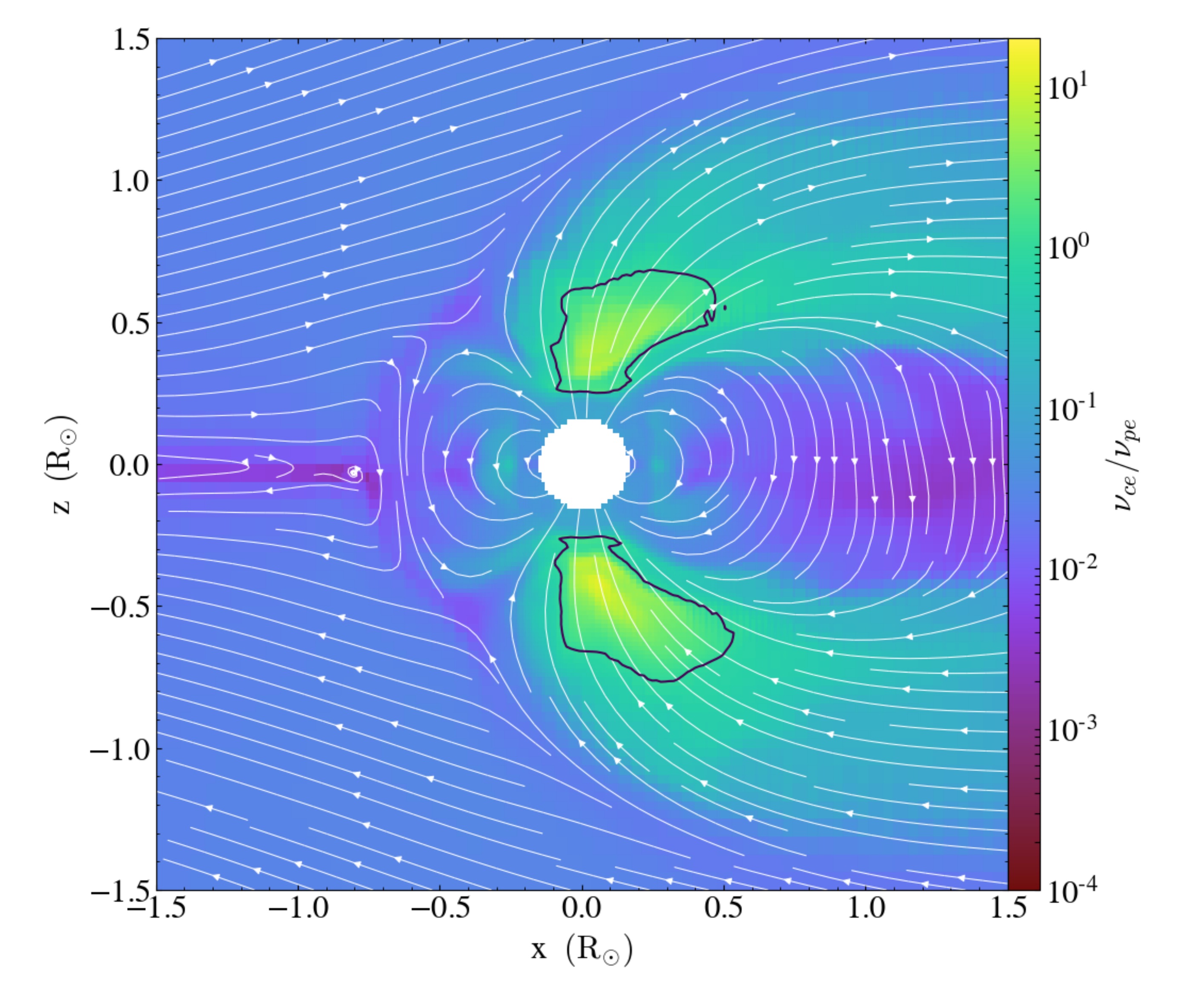}
\caption{Slice plots of the ratio $\nu_{\mathrm{ce}} / \nu_{\mathrm{pe}}$ with overplotted streamline indicating the magnetic field. Top: Model 2, the considerable amount of material evaporating from the surface completely inhibits the ECMI emission process. Bottom: Model 1, in this instance the amount of material in the vicinity of the planet is low enough to allow $\nu_{\mathrm{ce}} \ \nu_{\mathrm{pe}} > 1$. This region is enclosed by the black contour lines. $\nu_{\mathrm{ce}} \sim 0.021$ MHz at the point of highest $\nu_{\mathrm{ce}} / \nu_{\mathrm{pe}}$.
\label{fig:fcfp}}
\end{figure*}

Fig. \ref{fig:global} shows the final quasi-steady state solution. In both models the stellar wind has ripped open the magnetic field lines and the planetary outflow from the planet has been swept back. The ECMI emission efficiency is assessed by plotting the ratio in equation (\ref{eq:ECMI}) at every point in the proximity of the planet. Fig. \ref{fig:fcfp} shows a slice plot of this quantity for both planetary models. In the case of Model 2, the high mass-loss and therefore higher plasma density in the vicinity of the planet, increases the plasma frequency according to equation \ref{eq:pe}. This decreases the result of equation \ref{eq:ECMI} and is highly effective at inhibiting the EMCI emission process. In the case of Model 1, the planetary outflow density is an order of magnitude lower. This allows equation (\ref{eq:ECMI}) to exceed unity in the region above and below the planetary poles. The magnetic field strength is greatest here. The dipole topology of the planetary magnetic field acts to evacuate the material to the equatorial regions. Both of these factors lead to an increased ECMI efficiency and to the two volumes which can emit cyclotron radiation, which can be seen above and below the planet in Fig. \ref{fig:fcfp} bounded by the contour lines.

Emission generated in the polar regions of Model 1 has to escape the system in order to be detected by an observer. The emission frequency is constrained to be proportional to the magnitude of the magnetic field, according to equation (\ref{eq:ce}). For the models investigated here, the equatorial magnetic field is, $B_{\mathrm{eq}} =  1 \ \mathrm{G}$, the field strength at the poles is therefore $2 B_{\mathrm{eq}} = 2 \ \mathrm{G}$ (dipolar field). This means that $\nu_{\mathrm{ce}} \leq 5.6 \ \mathrm{MHz}$. In practice, this only occurs on the planets surface, a region which is subject to the highest outflow density and therefore highest plasma frequency, leading to a reduction in $\nu_{\mathrm{ce}} / \nu_{\mathrm{pe}}$. From Fig. \ref{fig:fcfp}, it can be see that the emission location is approximately $0.5 R_{\circ}$ above and below the planet. At this point the emission frequency is $\sim 0.021$ MHz. This value is however three orders of magnitude below the frequency cut off of the Earth's atmosphere. Therefore, if the emission is produced and can escape the system, then it would still be undetectable on earth, given the current capabilities of contemporary instruments.

\section{Conclusions}

The detection of radio emission from exoplanet systems would be an invaluable tool allowing a window on to rotational models, internal structure and evolution theories. Thus far there have been no confirmed observations of such emission from hot Jupiters, which are expected to be bright non-thermal radio sources.

This work has shown that the photoevaporation and resultant planetary wind which form around these bodies can act to increase the plasma density in the planetary magnetosphere, effectively blocking the ECMI emission process. This provides an explanation for the lack of detected radio emission. There is currently a more detailed paper in preparation which explained in more depth the models and results shown here. 

\section{Acknowledgments}

The authors acknowledge support from the Science and Technologies Facilities Research Council (STFC). 

Computations were performed using the University of Birmingham's BlueBEAR HPS service, which was purchased through HEFCE SRIF-3 funds. See http://www.bear.bham.ac.uk.

\bibliographystyle{an}
\bibliography{/Users/simon/Work/Reading/Reference_papers/Bibtex/library}

\label{lastpage}

\end{document}